# HIGHER ORDER CONTINUUM WAVE EQUATION CALIBRATED ON LATTICE DYNAMICS

Z. XU[1], R.C. PICU[2] AND J. FISH[1]

[1]*Department of Civil Engineering,*
[2]*Department of Mechanical, Aerospace and Nuclear Engineering,*
*Rensselaer Polytechnic Institute, Troy, NY 12180*

The classical approach to linking lattice dynamics properties to continuum equations of motion, the "method of long waves," is extended to include higher order terms. The additional terms account for non-local and non-linear effects. In the first part of the article, the derivation is made within the harmonic approximation for the perfect lattice response. Higher order terms are included in the continuum equation of motion to account for non-linear dispersion effects. Wave propagation coefficients as well as fourth order dispersion coefficients are obtained. In the second part, the lattice anharmonicity is considered and nonlinear macroscopic equations of motion are obtained within the local approximation. Both continuum solutions are particularized to the one-dimensional case and are compared with the lattice response in order to establish the accuracy of the approximation.

## 1. Introduction

Discrete and continuum systems are described by significantly different mathematical formulations. While the continuum behavior is typically represented by partial differential equations, the response of discrete media to the same perturbation is described by a set of discrete equations that represent the evolution of sets of degrees of freedom of the system. In many modern applications linking the two representations is crucial.

The discrete representation of ordered media such as crystal lattices, and that of disordered amorphous materials is no different. One identifies the relevant degrees of freedom and integrates the relevant equations of motion to trace the evolution of the system. Similarly, in the continuum sense, crystalline and amorphous structures are treated using the same formalism. However, when the discrete representation needs to be linked with the continuum, specificity becomes important. In this article we focus on the mechanical response of crystal lattices.

The discrete and periodic nature of the lattice imparts most of its outstanding properties. The discreteness leads to an intrinsically non-local nature of stress (which reflects in the equations of motion). The stress at given location is governed by the deformation within a whole neighborhood of that location. The discreteness also leads to a specific wave propagation behavior represented by a dispersion relationship leading to zero group velocity at the boundary of the Brillouin zone, and to several optical modes. Common local continuum models do not capture these features.

The periodic nature of the lattice determines other important properties. One of the most important is the non-convexity of the total potential energy function. A translation of a region of the crystal with respect to another by a multiple of the lattice vectors leads to recovering the perfect crystal and therefore the energy does not change as a result of this operation. Hence, the total energy function has multiple minima leading to many possible equilibrium configurations of the crystal. The periodicity also leads to multiple temporal scales, situation similar to that encountered in heterogeneous materials treated in the continuum sense.[1,2] Such continuum problems are solved within the homogenization theory using asymptotic analysis techniques. [3,4,5]

The lattice response is determined by the law describing the interatomic interactions. This function is non-parabolic and asymmetric with respect to the equilibrium position of two neighboring atoms. This leads to non-linear material behavior. The anharmonicity leads to notable macroscopic effects such as thermal expansion and finite lattice thermal conductivity.

Here we focus on lattice vibrations and the linkage of lattice dynamics with the continuum wave equation. The classical basis for this investigation is the work by Born and Huang[6] and later advances summarized by Venkataraman *et al*.[7] In the "method of long waves" detailed in[6], the lattice dynamics is treated within the harmonic approximation for the interatomic interactions. In this reference it is shown how this formulation for the discrete system converts into the classical wave equation in the continuum sense, and how the constants entering the continuum equation are related to the elastic constants of the lattice. Although the path to follow was sketched, in order to preserve the simplicity of the formulation, higher order effects were not incorporated in the classical solution .[6]

In this article we derive explicit expressions for the corrections to the continuum wave equation that account, in the first order, for a) non-local lattice effects, and b) non-linear material behavior. The derivation is made within the lattice dynamics theory and the correction terms are expressed in terms of quantities that may be evaluated from the discrete system.

## 2. Non-local Effects

### 2.1 The discrete system

A discrete system described by harmonic interatomic interactions is considered first. The goal here is to derive higher order wave propagation coefficients that capture the intrinsically non-local nature of these interactions.

The total potential energy of a deforming lattice, $\Phi$, may be expanded in series as a function of the local displacements, **u**. Using the notation in [6], $\Phi$ is expressed as:

$$\Phi = \Phi^{(0)} + \Phi^{(1)} + \Phi^{(2)} + \Phi^{(3)}... \qquad (1a)$$

$$\Phi^{(0)} = \Phi\left(\left\{\mathbf{x}\begin{pmatrix} l \\ j \end{pmatrix}\right\}\right). \tag{1b}$$

$$\Phi^{(1)} = \sum_{ljm} \left.\frac{\partial \Phi}{\partial u_m \begin{pmatrix} l \\ j \end{pmatrix}}\right|_0 u_m\begin{pmatrix} l \\ j \end{pmatrix} = \sum_{ljm} \phi_m\begin{pmatrix} l \\ j \end{pmatrix} u_m\begin{pmatrix} l \\ j \end{pmatrix}. \tag{1c}$$

$$\Phi^{(2)} = \frac{1}{2}\sum_{ljm}\sum_{l'j'n} \left.\frac{\partial^2 \Phi}{\partial u_m\begin{pmatrix} l \\ j \end{pmatrix}\partial u_n\begin{pmatrix} l' \\ j' \end{pmatrix}}\right|_0 u_m\begin{pmatrix} l \\ j \end{pmatrix}u_n\begin{pmatrix} l' \\ j' \end{pmatrix} = \frac{1}{2}\sum_{ljm}\sum_{l'j'n}\phi_{mn}\begin{pmatrix} l & l' \\ j & j' \end{pmatrix}u_m\begin{pmatrix} l \\ j \end{pmatrix}u_n\begin{pmatrix} l' \\ j' \end{pmatrix}. \tag{1d}$$

$$\Phi^{(3)} = \frac{1}{6}\sum_{ljm}\sum_{l'j'n}\sum_{l''j''w} \left.\frac{\partial^3 \Phi}{\partial u_m\begin{pmatrix} l \\ j \end{pmatrix}\partial u_n\begin{pmatrix} l' \\ j' \end{pmatrix}\partial u_w\begin{pmatrix} l'' \\ j'' \end{pmatrix}}\right|_0 u_m\begin{pmatrix} l \\ j \end{pmatrix}u_n\begin{pmatrix} l' \\ j' \end{pmatrix}u_w\begin{pmatrix} l'' \\ j'' \end{pmatrix} =$$

$$= \frac{1}{6}\sum_{ljm}\sum_{l'j'n}\sum_{l''j''w}\phi_{mnw}\begin{pmatrix} l & l' & l'' \\ j & j' & j'' \end{pmatrix}u_m\begin{pmatrix} l \\ j \end{pmatrix}u_n\begin{pmatrix} l' \\ j' \end{pmatrix}u_w\begin{pmatrix} l'' \\ j'' \end{pmatrix}. \tag{1e}$$

where, $(\bullet)|_0$ denotes the derivatives with respect to equilibrium configuration, $m$, $n$ and $o$ are Cartesian indices, and $j$ ($j'$ and $j''$) denotes the $j$th ($j'$th and $j''$th) atom in the $l$th cell. $\phi_{mn}\begin{pmatrix} l & l' \\ j & j' \end{pmatrix}$ denotes the force exerted on atom $\begin{pmatrix} l \\ k \end{pmatrix}$ in the $m$th direction due to a unit displacement of atom $\begin{pmatrix} l' \\ k' \end{pmatrix}$ in the $n$th direction. In this section in which the harmonic approximation is made, we neglect $\Phi^{(3)}$.

The equation of motion for atom $j$ in cell $l$ reads

$$m_j \ddot{u}_m\begin{pmatrix} l \\ j \end{pmatrix} + \sum_{l'j'n}\phi_{mn}\begin{pmatrix} l & l' \\ j & j' \end{pmatrix}u_n\begin{pmatrix} l' \\ j' \end{pmatrix} = 0. \tag{2}$$

for which a wavelike solution is sought in the form:

$$u_m\begin{pmatrix} l \\ j \end{pmatrix} = \frac{1}{\sqrt{m_j}}U_m(\mathbf{k})\exp\{i[\mathbf{k}\mathbf{x}(l)-\omega(\mathbf{k})t]\}. \tag{3}$$

Upon substitution in (2) the following classical equation is obtained:

$$D_{mnjj'}U_{mj}(\mathbf{k}) = \omega^2(\mathbf{k})U_{nj'}(\mathbf{k'}). \tag{4}$$

Here $U_m(\mathbf{k})$ is the wave amplitude, which is a function of the wave vector $\mathbf{k}$. $U_{mj}$ denotes the corresponding $j$th amplitude of $u_m\begin{pmatrix} l \\ j \end{pmatrix}$ and $D_{mnjj'}$ is the dynamic tensor with expression:

$$D_{mnjj'} = \frac{1}{\sqrt{m_j m_{j'}}} \sum_{l'} \phi_{mn}\begin{pmatrix} l & l' \\ j & j' \end{pmatrix} \exp\{i\mathbf{k}[\mathbf{x}(l') - \mathbf{x}(l)]\}. \tag{5}$$

Following further the classical path and limiting attention to the first Brillouin zone, eqn (4) may be rewritten as

$$D_{mnjj'}(\mathbf{k})B_{nj'}(\mathbf{k}) = A_{mn}(\mathbf{k})B_{nj}(\mathbf{k}). \tag{6}$$

which should be valid for any $j$ and $j'$. The dynamic matrix $A_{mn}(\mathbf{k})$ includes the information on the elastic properties and has the amplitude vectors $U_m(\mathbf{k})$ as its eigenvectors.

In order to render the notation more compact, we drop the specification of the atom and unit cell, $\begin{pmatrix} l \\ j \end{pmatrix}$, from all equations that follow. It is now useful to expand the dynamic matrix $A_{mn}(\mathbf{k})$ in a series of $k$ as:

$$A_{mn} = C^{(2)}_{mnpq}k_p k_q + C^{(3)}_{mnpqr}k_p k_q k_r + C^{(4)}_{mnpqrs}k_p k_q k_r k_s + \ldots\ldots \tag{7}$$

where the coefficients result by differentiation with respect to $k_m$, as usual:

$$C^{(2)}_{mnpq} = \frac{1}{2}\frac{\partial^2 A_{mn}}{\partial k_p \partial k_q}\bigg|_{\mathbf{k}=0},\ C^{(3)}_{mnpqr} = \frac{1}{6}\frac{\partial^3 A_{mn}}{\partial k_p \partial k_q \partial k_r}\bigg|_{\mathbf{k}=0},$$

$$C^{(4)}_{mnpqrs} = \frac{1}{24}\frac{\partial^4 A_{mn}}{\partial k_p \partial k_q \partial k_r \partial k_s}\bigg|_{\mathbf{k}=0}. \tag{8}$$

Similarly, the dynamic tensor $D_{mnjj'}$ and the amplitude matrix $B_{nj}$ may also be expanded in series with respect to $\mathbf{k}$ as:

$$D_{mnjj'} = D^{(0)}_{mnjj'} + D^{(1)}_{mnjj',w}k_w + D^{(2)}_{imnjj',pq}k_p k_q + D^{(3)}_{mnjj',pqr}k_p k_q k_r + D^{(4)}_{mnjj',pqrs}k_p k_q k_r k_s + \ldots\ldots$$
$$\tag{9}$$

$$B_{nj}(\mathbf{k}) = B^{(0)}_{nj} + B^{(1)}_{nj,w}k_w + B^{(2)}_{nj,pq}k_p k_q + B^{(3)}_{nj,pqr}k_p k_q k_r + B^{(4)}_{nj,pqrs}k_p k_q k_r k_s + \ldots\ldots \tag{10}$$

We note that all odd order terms in these expansions are imaginary.

*Symmetry properties and identities*

Each of the tensors on the right side of eqns (7) and (9), $\mathbf{C}^{(.)}$ and $\mathbf{D}^{(.)}$, follow symmetry relations which are described in [6]. Further details are provided in [8]. Furthermore, the following relations hold:

$$\sqrt{m_j} D^{(0)}_{mnjj'} = \sqrt{m_j} D^{(0)}_{nmj'j} = 0. \tag{11a}$$

$$\sqrt{m_j m_{j'}} D^{(1)}_{mnjj',w} = 0. \tag{11b}$$

$$\sqrt{m_j} D^{(1)}_{mnjj',w} = \sqrt{m_j} D^{(1)}_{mwjj',n}. \tag{11c}$$

Several other equations result by requiring that the wave propagation is insensitive to reversing the direction of the wave vector. Hence, the eigenvalues and eigenvectors of $A_{mn}(\mathbf{k})$ must have this symmetry property:

$$\omega_m(-\mathbf{k}) = \omega_m(\mathbf{k}). \tag{12a}$$

$$U_m(-\mathbf{k}) = U_m(\mathbf{k}). \tag{12b}$$

This reversal symmetry leads to the requirement that all odd order $\mathbf{C}^{(.)}$ tensors in the right side of eqn (7) must vanish. [8]

*Relations between higher order tensors $\mathbf{C}^{(.)}$ and $\mathbf{D}^{(.)}$*

Substituting eqns. (8), (9) and (10) into eqn (6) and identifying the coefficients of similar powers of $\mathbf{k}$ leads to a series of relations between tensors $\mathbf{C}^{(.)}$ and $\mathbf{D}^{(.)}$. The objective here is to establish a systematic procedure by which the higher order tensors $\mathbf{C}^{(.)}$ may be evaluated. These tensors enter the continuum equation of motion, as discussed in the next Section. The results up to the second order tensors were derived in the classical literature by a different procedure.[6] The contribution here consists in the method of derivation, which allows even higher order tensors in eqn (8) to be determined. For sake of simplicity, we limit the discussion to the fourth order member of the family, $\mathbf{C}^{(4)}$. The principal results are as follows:

$$C^{(2)}_{mnpq} = \frac{a_j a_{j'}}{a_w a_w} \left\{ D^{(1)}_{mujv,p} G^{(1)}_{unvj',q} + D^{(2)}_{mnjj',pq} \right\}. \tag{13a}$$

$$C^{(3)}_{mnpqr} = 0. \tag{13b}$$

$$C^{(4)}_{mnpqrs} = \frac{a_j a_{j'}}{a_w a_w} \left\{ \left( D^{(1)}_{mujv,s} G^{(3)}_{unvj',pqr} \right) + \left( \left[ D^{(2)}_{mujv,pq} - C^{(2)}_{mupq} \delta_{jv} \right] G^{(2)}_{unvj',rs} \right) + \left( D^{(3)}_{mujv,pqr} G^{(1)}_{unvj',s} \right) + D^{(4)}_{mnjj',pqrs} \right\}$$

$$\tag{13c}$$

where

$$G^{(1)}_{nuj'v,w} = -D^{(0)^{-1}}_{nmj'j} D^{(1)}_{mujv,w}. \tag{14a}$$

$$G^{(2)}_{nuj'v,pq} = -D^{(0)^{-1}}_{nmj'j}\left\{D^{(1)}_{mujv,p}G^{(1)}_{unvj',q} + \left[D^{(2)}_{mujv,pq} - C^{(2)}_{mupq}\delta_{jv}\right]\right\}. \tag{14b}$$

$$G^{(3)}_{nuj'v,pqr} = -D^{(0)^{-1}}_{nmj'j}\left\{\left(D^{(1)}_{mujv,r}G^{(2)}_{unvj',pq}\right) + \left(\left[D^{(2)}_{mujv,pq} - C^{(2)}_{mupq}\delta_{jv}\right]G^{(1)}_{unvj',r}\right) + \left(D^{(3)}_{mnjj',pqr}\right)\right\}. \tag{14c}$$

and $a_j = \sqrt{m_j}$, with $m_j$ being the mass of atom $j$. The detailed derivation is presented in Appendix A.

We note that although we restrict discussion to the fourth order term, the method presented here may be applied to higher order terms in the expansions (7) and (10). Certainly, retaining only the first terms leads to an approximation being introduced in the solution, which is only valid in the limit of long wavelengths. Therefore, although the present approach will improve upon the linear dispersion of the linear continuum (first order term only), phenomena occurring close to the boundary of the first Brillouin zone are not properly reproduced.

## 2.2 The continuum system

Let us now consider the continuum system. The Fourier transform of the macroscopic displacement field $\boldsymbol{u}(\boldsymbol{x},\mathrm{t})$ may be written formally as:

$$\boldsymbol{u}(\boldsymbol{x},\mathrm{t}) = \frac{1}{2\pi}\int_{-\infty}^{+\infty}\boldsymbol{F}(\boldsymbol{k})\begin{Bmatrix}e^{i(\boldsymbol{k}\boldsymbol{x}-\omega_1\mathrm{t})}\\ e^{i(\boldsymbol{k}\boldsymbol{x}-\omega_2\mathrm{t})}\\ e^{i(\boldsymbol{k}\boldsymbol{x}-\omega_3\mathrm{t})}\end{Bmatrix}d\boldsymbol{k}. \tag{15}$$

Here, $\mathbf{F}(\mathbf{k})$ is the $3\times 3$ transform functions matrix, and $\mathbf{k}$ is the wavevector in three dimensions. Unlike the discrete system case, integration is implemented here over the whole wavevector domain. For each wavevector, there are three corresponding frequencies, $\omega_1, \omega_2, \omega_3$. The equations that follow contain for simplicity only one frequency, however, the total displacement should be the sum of all three components. Then eqn (15) can be rewritten in the index form as:

$$u_m(\mathbf{x},t) = \frac{1}{2\pi}\int_{-\infty}^{+\infty}F_m(\mathbf{k})e^{i(k_w x_w - \omega t)}d\mathbf{k}. \tag{16}$$

From eqn (16), the time derivatives and spatial derivatives can be obtained directly, for example:

$$u_{m,t} = \frac{1}{2\pi}\int_{-\infty}^{+\infty}(-i\omega)F_m(\mathbf{k})e^{i(k_w x_w - \omega t)}d\mathbf{k}. \tag{17a}$$

$$u_{m,n} = \frac{1}{2\pi}\int_{-\infty}^{+\infty}ik_n F_m(\mathbf{k})e^{i(k_w x_w - \omega t)}d\mathbf{k}. \tag{17b}$$

while the strain and the strain gradients read

$$\left(u_{p,q}+u_{q,p}\right)=\frac{1}{2\pi}\int_{-\infty}^{+\infty}i\left(F_{p}k_{q}+F_{q}k_{p}\right)e^{i(k_{w}x_{w}-\omega t)}d\mathbf{k}\,. \tag{17c}$$

$$\left(u_{p,q}+u_{q,p}\right)_{,r}=\frac{1}{2\pi}\int_{-\infty}^{+\infty}\left(-k_{r}\right)\left(F_{p}k_{q}+F_{q}k_{p}\right)e^{i(k_{w}x_{w}-\omega t)}d\mathbf{k}\,. \tag{17d}$$

$$\left(u_{p,q}+u_{q,p}\right)_{,rst}=\frac{1}{2\pi}\int_{-\infty}^{+\infty}k_{r}k_{s}k_{t}\left(F_{p}k_{q}+F_{q}k_{p}\right)e^{i(k_{w}x_{w}-\omega t)}d\mathbf{k}\,. \tag{17e}$$

The crucial step is now to define the eigenfrequencies and the eigenvectors as common variables between the atomistic discrete level and the continuum representation [9]. This allows the incorporation of information from lattice dynamics into the continuum equations of motion. Consequently, the discrete dispersion equation is replaced in the continuum relationship to get:

$$\frac{1}{2\pi}\int_{-\infty}^{+\infty}\left\{A_{mn}(\mathbf{k})-\omega^{2}(\mathbf{k})\delta_{mn}\right\}F_{n}e^{i(\mathbf{k}\mathbf{x}-\omega t)}d\mathbf{k}=0 \text{ for any } \mathbf{x},\mathbf{t}\,, \tag{18}$$

where $\mathbf{F}$ of the continuum formulation of eqn (16) is required to be identical to the eigenvector $\mathbf{U}$ of $A$.

The dynamic matrix in eqn (18), $A_{mn}(\mathbf{k})$, is expanded in series as in eqn (7) and only terms up to the fourth order are preserved. After rearrangement, eqn (18) reads:

$$u_{m,tt}=\frac{1}{2\pi}\int_{-\infty}^{+\infty}\left\{C_{mnpq}^{(2)}k_{p}k_{q}+C_{mnpqr}^{(3)}k_{p}k_{q}k_{r}+C_{mnpqrs}^{(4)}k_{p}k_{q}k_{r}k_{s}\right\}F_{n}e^{i(\mathbf{k}\mathbf{x}-\omega t)}d\mathbf{k}\,.\tag{19}$$

Eqn (19) may be simplified using the spatial derivatives of the displacement field as in eqn (17). The macroscopic equation of motion (up to the fourth order) becomes:

$$u_{m,tt}-C_{mnpq}^{(2)}u_{n,pq}=iC_{mnpqr}^{(3)}u_{n,pqr}+C_{mnpqrs}^{(4)}u_{n,pqrs}\,. \tag{20}$$

which, considering that the reversal symmetry needs to be satisfied (and $C^{(3)}=\mathbf{0}$), simplifies to:

$$u_{m,tt}-C_{mnpq}^{(2)}u_{n,pq}=C_{mnpqrs}^{(4)}u_{n,pqrs}\,. \tag{21}$$

It is useful to rewrite eqn (21) in terms of strain (small strain formulation). To this end we use the procedure underlined in [6] to get

$$u_{m,tt}-L_{mnpq}^{2}\varepsilon_{pq,n}=L_{mnpqrs}^{4}\varepsilon_{pq,nrs}\,. \tag{22}$$

where $L_{mnpq}^{2}$ is the second order elastic constants tensor. According to [6], the second order $L$ may be obtained from the second order tensor $C$ with the relation $L^2_{mnpq} = C^{(2)}_{mpnq} + C^{(2)}_{pnmq} - C^{(2)}_{pqmn}$. This may be generalized to the fourth order tensors and reads

$$L^4_{mnpqrs} = C^{(4)}_{mpnqrs} + C^{(4)}_{pnmqrs} - C^{(4)}_{pqmnrs}. \quad (23)$$

The additional fourth order term in eqn (22) represents the effect of the non-local nature of the interactions in the lattice. The continuum wave equation captures the dispersion effect due to non-locality only in an approximate way since only terms up to the fourth order in the expansion are considered. The accuracy of this approximation is evaluated in the next Section on a particular case.

## 2.3 The one-dimensional case

For the purpose of illustration, the 3D formulation presented above is reduced here to 1D. We consider a one-dimensional atomic chain consisting of atoms with different masses, $m_1$ and $m_2$, being connected by linear springs with different stiffness, $\chi_1$ $\chi_2$. The mass points (atoms) are equally separated at a distance $a$. For this system, the dynamic tensor $D$, eqn (5), reduces to:

$$D_{11jj'} = \begin{bmatrix} \dfrac{(\chi_1 + \chi_2)}{m_1} & -\dfrac{\chi_1 e^{-ika} + \chi_2 e^{ika}}{\sqrt{m_1 m_2}} \\ -\dfrac{\chi_1 e^{ika} + \chi_2 e^{-ika}}{\sqrt{m_1 m_2}} & \dfrac{(\chi_1 + \chi_2)}{m_2} \end{bmatrix} \quad j, j' = 1,2. \quad (24)$$

Then, using eqns (13) it results that the second and fourth order $C$ tensors contain only one nonzero component, namely

$$C^{(2)}_{1111} = a\sqrt{\dfrac{\overline{\chi}}{\overline{m}}}. \quad (25a)$$

$$C^{(4)}_{111111} = -\dfrac{a^4}{3}\dfrac{\overline{\chi}}{\overline{m}}\left[1 - \dfrac{6\overline{\chi}}{\chi_1 + \chi_2} \cdot \dfrac{m_1 m_2}{(m_1 + m_2)^2}\right]. \quad (25b)$$

where the equivalent stiffness and mass are given by the usual relations

$$\overline{\chi} = \dfrac{2\chi_1 \chi_2}{\chi_1 + \chi_2} \qquad \overline{m} = \dfrac{m_1 + m_2}{2}. \quad (26)$$

With this, the corresponding macroscopic equation of motion, eqn (21), becomes:

$$u_{tt} - C^{(2)}_{1111} u_{xx} = C^{(4)}_{111111} u_{xxxx}. \quad (27)$$

Within the local harmonic approximation, the right hand side term in eqn (27) disappears and the one-dimensional wave equation is recovered.

In order to evaluate the improvement of the continuum solution obtained by the addition of the higher order term, a numerical example was considered. A chain of 200 atoms of two species having a mass ratio of 10 and connected by harmonic springs of different stiffnesses (stiffness ratio 100) is perturbed at time zero by a displacement described by the Gaussian $u(x) = 0.01 \times \exp\left(-\left(\frac{x-20}{1.2}\right)^2\right)$. The system is evolved in time a) by tracing the trajectory of each atom (the lattice dynamics solution), b) by integrating the local wave equation (eqn (27)) without the higher order term (Solution 1), and c) by integrating the corrected eqn (27) (Solution 2). The displacement profile at time t=10 as predicted by the three methods is shown in Fig. 1. The wave has propagated from the initial perturbation centered in the middle of the domain and has reflected on the left and right boundaries of the model. The left boundary is fixed, while the right one is free. It is seen that considering the non-local correction (Solution 2 in Fig. 1) improves upon the local solution (Solution 1). Better agreement with the lattice dynamics result may be obtained by considering higher order correction terms in eqn. (27).

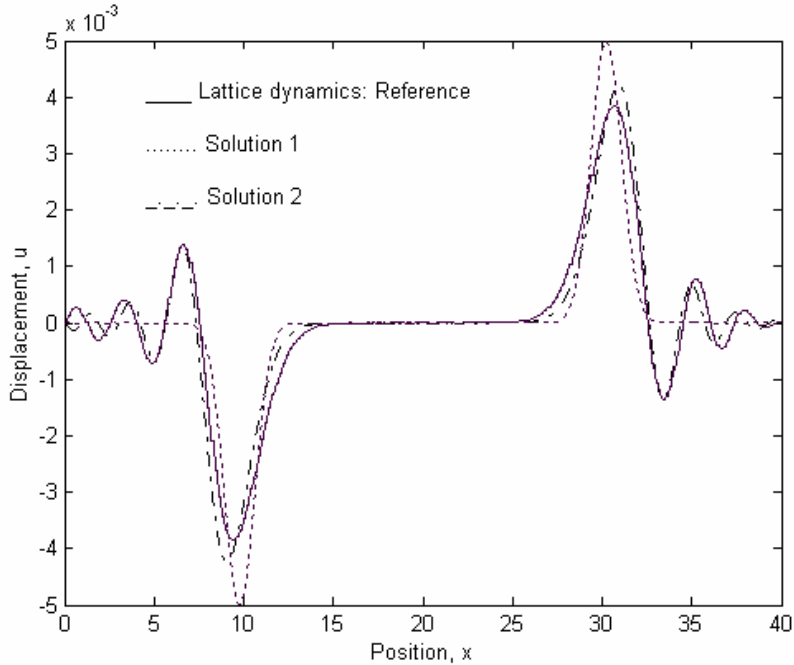

Figure 1. Comparison of solutions for the non-local harmonic one-dimensional problem. The reference solution (continuous line) was obtained by lattice dynamics. Solution 1 results by integrating the continuum wave equation (27) in its local approximation, without the fourth order term, while Solution 2 is the non-local result.

## 3. Anharmonic Effects

### 3.1 The discrete system

The approach described in Section 2 may be used unchanged to incorporate non-linear effects that originate in the anharmonicity of the lattice in the continuum wave equation. Although the result, including the role of the higher order elastic constants tensors, is known,[10,11] we review it here based on the present formulation.

The anharmonic behavior of the lattice is an important detail of the physics; it leads to coupling between vibration modes of the lattice, coupling that controls the phonon mean free path and the reduction of the lattice thermal conductivity with temperature, as well as certain absorption phenomena. In order to include anharmonic effects, the third and higher order terms in the expansion of the lattice potential energy need to be considered. For simplicity, the third term only ($\Phi^{(3)}$) is typically considered. The equation of motion for atom $j$ in lattice cell $l$ (eqn (2)) becomes:

$$m_j \ddot{u}_m \binom{l}{j} + \sum_{l'j'n}\left\{\phi_{mn}\binom{l\ l'}{j\ j'} + \sum_{l''j''w}\phi_{mnw}\binom{l\ l'\ l''}{j\ j'\ j''}u_w\binom{l''}{j''}\right\}u_n\binom{l'}{j'} = 0. \quad (28)$$

With the notation in [6], eqn (28) may be rewritten in the form

$$m_j \ddot{u}_m \binom{l}{j} + \sum_{l'j'n}\left\{\phi_{mn}\binom{l\ l'}{j\ j'} + \sum_{l''j''w}\phi_{mnw}\binom{l\ l'\ l''}{j\ j'\ j''}u_w\binom{l''-l}{j''\ j}\right\}u_n\binom{l'}{j'} = 0. \quad (29)$$

where, $u_w\binom{l''-l}{j''\ j} = u_w\binom{l''}{j''} - u_w\binom{l}{j}$. This expresses the fact that, due to the periodicity of the lattice (invariance with respect to translations by multiples of lattice vectors), the first order coefficients are independent of the cell index $l$, while higher order coefficients must depend on the relative distance between cells only, or conversely, on the relative cell index $l - l''$. Therefore, it is more natural to replace the relative displacement $u_w\binom{l''-l}{j''\ j}$ in eqn (29) by the Lagrangian strain, $\varepsilon_{wp}$ (assume small strains). The two quantities are related by

$$u_w\binom{l''-l}{j''\ j} = \varepsilon_{wp} x_p\binom{l''-l}{j''\ j}. \quad (30)$$

The solution of eqn (29) is sought in the form of a superposition of wavelike functions of the form (3). After substitution in (29), the following system of equations results:

$$D^A_{mnjj''}U_{mj} = \omega^2(\mathbf{k})U_{nj'}. \quad (31)$$

where $D^{A}_{mnjj'}$ is the dynamic tensor that includes anharmonic effects. This tensor may be decomposed (due to the summation in the second terms in eqn (28)) in a harmonic component and a strain-dependent correction:

$$D^{A}_{mnjj'} = D^{H}_{mnjj'} + D^{T}_{mnpqjj'} \cdot \varepsilon_{pq}. \tag{32a}$$

The harmonic term is identical to that in eqn (5). The correction tensor is given by

$$D^{T}_{mnpqjj'} = \frac{1}{\sqrt{m_j m_{j'}}} \sum_{l'} \sum_{l''j''} \left\{ \phi_{mnp} \begin{pmatrix} l & l' & l'' \\ j & j' & j'' \end{pmatrix} x_q \begin{pmatrix} l''-l \\ j'' & j \end{pmatrix} \right\} \exp\{i\mathbf{k}[\mathbf{x}(l') - \mathbf{x}(l)]\}. \tag{32b}$$

The derivation follows the path detailed in Section 2. The $D$ tensors are expanded in a series of $\mathbf{k}$. Due to the linear superposition of the terms in the expansion, each coefficient may be decomposed as in eqn (32a):

$$D^{A(0)}_{mnjj'} = D^{H(0)}_{mnjj'} + D^{T(0)}_{mnrsjj'} \varepsilon_{rs}. \tag{33a}$$

$$D^{A(1)}_{mnjj',w} = D^{H(1)}_{mnjj',w} + D^{T(1)}_{mnrsjj',w} \varepsilon_{rs}. \tag{33b}$$

$$D^{A(2)}_{mnjj',pq} = D^{H(2)}_{mnjj',pq} + D^{T(2)}_{mnrsjj',pq} \varepsilon_{rs}. \tag{33c}$$

$G$ tensors follow the same rule. We note that only terms up to the second order are considered here since non-local effects are neglected. The relevant tensors were derived as follows:

$$C^{A(2)}_{mnpq} = C^{H(2)}_{mnpq} + C^{T(2)}_{mnpqrs} \varepsilon_{rs}. \tag{34a}$$

$$C^{H(2)}_{mnpq} = \frac{a_j a_{j'}}{a_w a_w} \left\{ D^{H(1)}_{mujv,p} G^{H(1)}_{unvj',q} + D^{H(2)}_{mnjj',pq} \right\}. \tag{34b}$$

$$C^{T(2)}_{mnpqrs} = \frac{a_j a_{j'}}{a_w a_w} \left\{ D^{H(1)}_{mujv,p} G^{T(1)}_{unvj'rs,q} + D^{T(1)}_{mujvrs,p} G^{H(1)}_{unvj',q} + D^{T(2)}_{mnrsjj',pq} \right\}. \tag{34c}$$

Both the harmonic and the correction tensors $G$ in these expressions are obtained as:

$$G^{A(1)}_{nuj'v,w} = G^{H(1)}_{nuj'v,w} + G^{T(1)}_{nuj'vrs,w} \varepsilon_{rs} \tag{35a}$$

$$G^{H(1)}_{nuj'v,w} = -D^{H(0)^{-1}}_{nmj'j} D^{H(1)}_{mujv,w} \tag{35b}$$

$$G^{T(1)}_{nuj'vrs,w} = -\left\{ D^{T(0)^{-1}}_{nmj'jrs} D^{H(1)}_{mujv,w} + D^{H(0)^{-1}}_{nmj'j} D^{T(1)}_{mujvrs,w} \right\} \tag{35c}$$

*3.2 The continuum system*

The equation of motion for the continuum in the local approximation reads:

$$u_{m,tt} - C^{A(2)}_{mnpq} u_{n,pq} = 0. \tag{36}$$

where the tensor $C_{mnpq}^{A(2)}$ includes non-linear effects as discussed above. The calibration of the continuum model to the lattice response is performed by simply replacing eqn (34a) in (36). The resulting equation may be written in term of strains (similar to the transformation of eqn (21) in (22)) as:

$$u_{m,tt} - \left(L_{mnpq}^{H2} + L_{mnpqrs}^{T2}\varepsilon_{rs}\right)\varepsilon_{pq,n} = 0. \tag{37}$$

$L_{mnpq}^{H2}$ and $L_{mnpqrs}^{T2}$ are the second order and the third order elastic constants tensors, respectively [11]. According to [6], the relation between the elastic constants and the corresponding wave propagation coefficients is:

$$L_{mnpq}^{H2} = C_{mpnq}^{H(2)} + C_{pnmq}^{H(2)} - C_{pqmn}^{H(2)} \tag{38a}$$

for the harmonic component and

$$L_{mnpqrs}^{T2} = C_{mpnqrs}^{T(2)} + C_{pnmqrs}^{T(2)} - C_{pqmnrs}^{T(2)} \tag{38b}$$

for the third order elastic constants.

### *3.3 The one-dimensional case*

The particularization of the 3D formulation described above to the 1D case is presented next. For this purpose, we consider a chain of atoms of same mass, *m*, linked by nonlinear springs. The spring stiffness is taken to be:

$$\chi = \chi_1 + \chi_2 x \tag{39}$$

$\chi_1$ and $\chi_2$ are elastic constants and *x* is the elongation. Substituting in eqns (5) and (32b), and with eqn (32a), the dynamic tensors in one dimension result in the form:

$$D_{1111}^{H(0)} = \frac{\chi_1}{m}\{2 - 2\cos(ka)\} \tag{40a}$$

$$D_{111111}^{T(0)} = \frac{2\chi_2}{m}a\{2 - 2\cos(ka)\} \tag{40b}$$

The wave propagation coefficients tensors result from eqns (34b) and (34c):

$$C_{1111}^{H(2)} = \frac{\chi_1}{m}a^2 \tag{41a}$$

$$C_{111111}^{T(2)} = 2\frac{\chi_2}{m}a^3 \tag{41b}$$

With the lattice response in the form (41), the corresponding continuum equation may be directly written based on eqn (36) as:

$$u_{tt} - \frac{a^2}{m}\left(\chi_1 + 2\chi_2 a u_{,x}\right) u_{,xx} = 0 \qquad (42)$$

This equation is identical to that derived in [12] for a chain of grains interacting by Hertz law. The derivation in [12] is limited to one dimensional problems only.

**4. Conclusions**

Continuum equations of motion incorporating corrections terms that account for non-local and non-linear effects (up to first order) have been derived. The new terms are calibrated on the lattice response and derived from considerations related to the discrete system.

**Appendix A**

In this Appendix, we present the derivation of the second order wave propagation coefficients $C^{(2)}_{mnpq}$ and of the high order dispersion coefficients $C^{(4)}_{mnpqrs}$. Further detail may be found in [8].

*Zero order*

By equating the coefficients of terms in the series expansion having zero power of $k$ yields:

$$D^{(0)}_{mnjj'} B^{(0)}_{nj'} = 0. \tag{A1a}$$

$$D^{(0)}_{mnjj'} B^{(0)}_{mj} = 0. \tag{A1b}$$

In light of eqn (11a), eqn (A1a) has the following solution:

$$B^{(0)}_{nj'} = a_{j'} U_n \qquad a_{j'} = \sqrt{m_{j'}}. \tag{A2}$$

where $U$ is an arbitrary vector in space which is determined in the analysis of the second order term. Equation (11a) shows that $D^{(0)}_{mnjj'}$ does not have an inverse, however, its sub-tensor can be inverted. This leads to

$$D^{S(0)}_{mnjj'} = D^{(0)}_{mnjj'}. \tag{A3a}$$

$$D^{S(0)^{-1}}_{nmj'j} \cdot D^{S(0)}_{mujv} = \delta_{nu} \delta_{j'v}. \tag{A3b}$$

where $S$ denotes a sub-tensor with $j$ and $j'$ range from 1 to ($N$-1).

*First order*

By equating of coefficients at the first order of $k$ yields:

$$D^{(0)}_{mnjj'} B^{(1)}_{nj',w} = -D^{(1)}_{mnjj',w} B^{(0)}_{nj'}. \tag{A4a}$$

$$D^{(0)}_{mnjj'} B^{(1)}_{mj,w} = D^{(1)}_{mnjj',w} B^{(0)}_{mj}. \tag{A4b}$$

$$D^{(1)}_{mnjj',w} B^{(0)}_{nj'} B^{(0)}_{mj} = 0. \tag{A4c}$$

Eqns. (A4a) and (A4b) are solvable only if eqn (A4c) is fulfilled. However, eqn (A4c) is satisfied due to the identity in eqn (11b) which shows that $D^{(1)}_{mnjj',w}$ is antisymmetric. Then, eqn (A4a) has a solution of the form:

$$B^{(1)}_{nj',w} = B^{(1)(p)}_{nj',w} + n_1 \cdot B^{(0)}_{nj'}. \tag{A5}$$

where $n_1$ is an arbitrary number. $B^{(1)(p)}_{nj',w}$ denotes the particular solution of eqn (A4a) and can be determined from:

$$B_{nj',w}^{(1)(p)} = G_{nuj'v,w}^{(1)} B_{uv}^{(0)}. \tag{A6}$$

Since the tensor $G_{nuj'v,w}^{(1)}$ should satisfy:

$$\left\{ D_{mnjj'}^{(0)} G_{nuj'v,w}^{(1)} + D_{mujv,w}^{(1)} \right\} B_{uv}^{(0)} = 0 \tag{A7}$$

$G_{nuj'v,w}^{(1)}$ can be determined from the following equations:

$$D_{mnjj'}^{(0)} G_{nuj'v,w}^{(1)} = -D_{mujv,w}^{(1)} \tag{A8a}$$

$$G_{nuj'v,w}^{(1)} = -D_{nmj'j}^{(0)^{-1}} D_{mujv,w}^{(1)} \tag{A8b}$$

where, $\quad D_{mnjj'}^{(0)^{-1}} = D_{mnjj'}^{S(0)^{-1}} \quad j \le N-1 \text{ and } j' \le N-1 \tag{A8c}$

$$D_{mnjj'}^{(0)^{-1}} = 0 \quad j = N \text{ or } j' = N \tag{A8d}$$

<u>Remark</u>: It can be verified that these tensors have the following properties.

$$D_{mnjj'}^{(0)^{-1}} = D_{nmj'j}^{(0)^{-1}} \tag{A9}$$

$$G_{mnjj'}^{(1)} = -G_{nmj'j}^{(1)} \tag{A10}$$

*Second order*

By equating of coefficients at the second order of *k* yields:

$$D_{mnjj'}^{(0)} B_{nj',pq}^{(2)} = -\left\{ \left( D_{mnjj',p}^{(1)} B_{nj',q}^{(1)} \right)_{\underline{pq}} + \left[ D_{mnjj',pq}^{(2)} - C_{mnpq}^{(2)} \delta_{jj'} \right] B_{nj'}^{(0)} \right\} \tag{A11a}$$

$$\left( D_{mnjj',p}^{(1)} B_{nj',q}^{(1)} \right)_{\underline{pq}} B_{mj}^{(0)} + \left[ D_{mnjj',pq}^{(2)} - C_{mnpq}^{(2)} \delta_{jj'} \right] B_{nj'}^{(0)} B_{mj}^{(0)} = 0 \tag{A11b}$$

where $(\bullet)_{\underline{pq}}$ denotes an operator on $(\bullet)$ that insures that $(\bullet)$ is symmetric over underlined subscripts. Equation (A11b) is the solvable condition for eqn (A11a). By substitution of the solution of eqn (A5) into equation (A11b), changing the subscripts and using equation (A4c) to eliminate the homogeneous solution part, we obtain:

$$\left( D_{mujv,p}^{(1)} G_{unvj',q}^{(1)} \right)_{\underline{pq}} B_{nj'}^{(0)} B_{mj}^{(0)} + \left[ D_{mnjj',pq}^{(2)} - C_{mnpq}^{(2)} \delta_{jj'} \right] B_{nj'}^{(0)} B_{mj}^{(0)} = 0. \tag{A12a}$$

We denote $\quad F_{mnjj'pq}^{21} = D_{mujv,p}^{(1)} G_{unvj',q}^{(1)}$

$$F_{mnjj'pq}^{22} = D_{mnjj',pq}^{(2)} - C_{mnpq}^{(2)} \delta_{jj'}. \tag{A12b}$$

and show the following symmetry properties for tensor *F*:

$$F_{mnjj'pq}^{21} = F_{nmj'jpq}^{21} = F_{mnjj'qp}^{21}. \tag{A13}$$

The solution for $B_{nj',pq}^{(2)}$ is:

$$D^{(0)}_{mnjj'}B^{(2)}_{nj',pq} = -\left\{F^{21}_{mnjj'pq} + F^{22}_{mnjj'pq}\right\}B^{(0)}_{nj'} - n_1\left(D^{(1)}_{mnjj',p}\right)_{\underline{pq}}B^{(0)}_{nj'}. \quad \text{(A14a)}$$

$$B^{(2)}_{nj',pq} = B^{(2)(p)}_{nj',pq} + n_1\left(B^{(1)(p)}_{nj',p}\right)_{\underline{pq}} + n_2 B^{(0)}_{nj'}. \quad \text{(A14b)}$$

$$B^{(2)(p)}_{nj',pq} = G^{(2)}_{nuj'v,pq}B^{(0)}_{uv}. \quad \text{(A14c)}$$

$$G^{(2)}_{nuj'v,pq} = -D^{(0)^{-1}}_{nmj'j}\left\{F^{21}_{mujvpq} + \left[D^{(2)}_{mujv,pq} - C^{(2)}_{mupq}\delta_{jv}\right]\right\}. \quad \text{(A14d)}$$

Substitution of the solution (A2) back into equation (A11b) and after simplifying using equation (A4c) the solution for the second order tensor $C^{(2)}_{mnpq}$ results as:

$$C^{(2)}_{mnpq} = \frac{a_j a_{j'}}{a_w a_w}\left\{F^{21}_{mnjj'pq} + D^{(2)}_{mnjj',pq}\right\}. \quad \text{(A15)}$$

This result for the second order wave propagation coefficients is identical to that obtained in [6] using the method of long waves.

*Third order*

Considering the coefficients of the third order in *k* leads to:

$$D^{(0)}_{mnjj'}B^{(3)}_{nj',pqr} = -\left\{\left(D^{(1)}_{mnjj',r}B^{(2)}_{nj',pq}\right)_{\underline{pqr}} + \left(\left[D^{(2)}_{mnjj',pq} - C^{(2)}_{mnpq}\delta_{jj'}\right]B^{(1)}_{nj',r}\right)_{\underline{pqr}} + \left(D^{(3)}_{mnjj',pqr} - C^{(3)}_{mnpqr}\delta_{jj'}\right)B^{(0)}_{nj'}\right\}$$

(A16a)

$$\left(D^{(1)}_{mnjj',r}B^{(2)}_{nj',pq}\right)_{\underline{pqr}}B^{(0)}_{mj} + \left(\left[D^{(2)}_{mnjj',pq} - C^{(2)}_{mnpq}\delta_{jj'}\right]B^{(1)}_{nj',r}\right)_{\underline{pqr}}B^{(0)}_{mj} + \left(D^{(3)}_{mnjj',pqr} - C^{(3)}_{mnpqr}\delta_{jj'}\right)B^{(0)}_{nj'}B^{(0)}_{mj} = 0$$

(A16b)

Eqn (A16b) is the condition for equation (A16a) to be solvable. Substitution of the solution of (A14b) into equations (A16a) and (A16b), and using equations (A12a) and (A4c) to eliminate the homogeneous parts results into:

$$\left(D^{(1)}_{mujv,r}G^{(2)}_{unvj',pq}\right)_{\underline{pqr}}B^{(0)}_{nj'}B^{(0)}_{mj} + \left\{\left[D^{(2)}_{mnjj',pq} - C^{(2)}_{mnpq}\delta_{jj'}\right]G^{(1)}_{unvj',r}\right\}_{\underline{pqr}}B^{(0)}_{nj'}B^{(0)}_{mj} + \left(D^{(3)}_{mnjj',pqr} - C^{(3)}_{mnpqr}\delta_{jj'}\right)B^{(0)}_{nj'}B^{(0)}_{mj} = 0$$

(A17)

and

$$D^{(0)}_{mnjj'}B^{(3)}_{nj',pqr} = -\left\{\left(D^{(1)}_{mujv,r}G^{(2)}_{unvj',pq}\right)_{\underline{pqr}} + \left(\left[D^{(2)}_{mujv,pq} - C^{(2)}_{mupq}\delta_{jv}\right]G^{(1)}_{unvj',r}\right)_{\underline{pqr}} + \left(D^{(3)}_{mnjj',pqr}\right)\right\}B^{(0)}_{nj'} -$$

$$n_1 \left( D^{(1)}_{mujv,p} G^{(1)}_{unvj',q} + D^{(2)}_{mnjj',pq} - C^{(2)}_{mnpq} \delta_{jj'} \right)_{\underline{pqr}} B^{(0)}_{nj'} - n_2 \left( D^{(1)}_{mnjj',r} \right)_{\underline{pqr}} B^{(0)}_{nj'} \quad \text{(A18)}$$

Note that under the condition of reversal symmetry, $C^{(3)}_{mnpqr}$ equals **0** [8] and eqn. (A17) is automatically satisfied.

The solution for $B^{(3)}_{nj',pq}$ is:

$$B^{(3)}_{nj',pqr} = B^{(3)(p)}_{nj',pqr} + n_1 \left( B^{(2)(p)}_{nj',pq} \right)_{\underline{pqr}} + n_2 \left( B^{(1)(p)}_{nj',p} \right)_{\underline{pqr}} + n_3 B^{(0)}_{nj'}. \quad \text{(A19a)}$$

$$B^{(3)(p)}_{nj',pqr} = G^{(3)}_{nuj'v,pqr} B^{(0)}_{uv}. \quad \text{(A19b)}$$

$$G^{(3)}_{nuj'v,pqr} = -D^{(0)-1}_{nmj'j} \left\{ \left( D^{(1)}_{mxjy,r} G^{(2)}_{xuyv,pq} \right)_{\underline{pqr}} + \left( \left[ D^{(2)}_{mxjy,pq} - C^{(2)}_{mxpq} \delta_{jy} \right] G^{(1)}_{xuyv,r} \right)_{\underline{pqr}} + \left( D^{(3)}_{mujv,pqr} \right) \right\}$$

(A20)

*Fourth order*

Considering the coefficients of the fourth order in *k* leads to:

$$D^{(0)}_{mnjj'} B^{(4)}_{nj',pqrs} = -\left\{ \left( D^{(1)}_{mnjj',r} B^{(3)}_{nj',pqr} \right)_{\underline{pqrs}} + \left( \left[ D^{(2)}_{mnjj',pq} - C^{(2)}_{mnpq} \delta_{jj'} \right] B^{(2)}_{nj',rs} \right)_{\underline{pqrs}} + \right\} \left\{ + \left( \left[ D^{(3)}_{mnjj',pqr} - C^{(3)}_{mnpqr} \delta_{jj'} \right] B^{(1)}_{nj',s} \right)_{\underline{pqrs}} + \left[ D \right. \right.$$

$$\left( D^{(1)}_{mnjj',r} B^{(3)}_{nj',pqr} \right)_{\underline{pqrs}} B^{(0)}_{mj} + \left( \left[ D^{(2)}_{mnjj',pq} - C^{(2)}_{mnpq} \delta_{jj'} \right] B^{(2)}_{nj',rs} \right)_{\underline{pqrs}} B^{(0)}_{mj} + \left( \left[ D^{(3)}_{mnjj',pqr} - C^{(3)}_{mnpqr} \delta_{jj'} \right] B^{(1)}_{nj',s} \right)_{\underline{pqrs}} B^{(0)}_{mj} +$$

$$\left[ D^{(4)}_{mnjj',pqrs} - C^{(4)}_{mnpqrs} \delta_{jj'} \right] B^{(0)}_{nj'} B^{(0)}_{mj} = 0 \quad \text{(A21b)}$$

As shown above, the homogeneous part always vanishes. From now on, only the particular solution will be considered and the odd *C* tensors should be **0**. Substitution of the solution of (A19b) into (A21a) and (A21b) and applying symmetric properties yields:

$$D^{(0)}_{mnjj'} B^{(4)}_{nj',pqrs} = -\left\{ \left( D^{(1)}_{mujv,s} G^{(3)}_{unvj',pqr} \right)_{\underline{pqrs}} + \left( \left[ D^{(2)}_{mujv,pq} - C^{(2)}_{mupq} \delta_{jv} \right] G^{(2)}_{unvj',rs} \right)_{\underline{pqrs}} + \right\}$$

$$\left\{ \left( D^{(3)}_{mujv,pqr} G^{(1)}_{unvj',s} \right)_{\underline{pqrs}} + \left( D^{(4)}_{mnjj',pqrs} - C^{(4)}_{mnpqrs} \delta_{jj'} \right) \right\} B^{(0)}_{nj'} \quad \text{(A22)}$$

$$C^{(4)}_{mnpqrs} = \frac{a_j a_{j'}}{a_w a_w} \left\{ \left( D^{(1)}_{mujv,s} G^{(3)}_{unvj',pqr} \right) + \left( \left[ D^{(2)}_{mujv,pq} - C^{(2)}_{mupq} \delta_{jv} \right] G^{(2)}_{unvj',rs} \right) + \left( D^{(3)}_{mujv,pqr} G^{(1)}_{unvj',s} \right) + D^{(4)}_{mnjj',pqrs} \right\}$$

(A23)